\documentclass[sigconf,10pt,nonacm]{acmart}

\makeatletter
\renewcommand{\@authorfont}{\normalsize}
\renewcommand{\@affiliationfont}{\small}
\makeatother

\usepackage{booktabs}
\usepackage{tabularx}
\usepackage{array}
\usepackage{multirow}
\usepackage{amsmath}
\usepackage{graphicx}
\usepackage{placeins}
\usepackage{xcolor}
\usepackage{seqsplit}

\newcommand{\bench}{\textsc{NetArtifactBench}}
\newcolumntype{Y}{>{\raggedright\arraybackslash}X}
\newcommand{\contractplain}[1]{#1}
\newcommand{\contractlow}[1]{#1}
\newcommand{\contractmid}[1]{#1}
\newcommand{\contracttop}[1]{\textbf{#1}}
\newcommand{\metricmeanstd}[2]{\ensuremath{#1\!\pm\!#2}}
\newcommand{\metricstd}[1]{\ensuremath{\;\pm\;#1}}
\definecolor{findingborder}{gray}{0.55}
\definecolor{findingbackground}{gray}{0.96}
\newcounter{finding}
\newcommand{\findingbox}[1]{%
  \par\addvspace{0.5\baselineskip}
  \refstepcounter{finding}%
  \noindent\begingroup
  \setlength{\fboxrule}{0.5pt}%
  \setlength{\fboxsep}{7pt}%
  \fcolorbox{findingborder}{findingbackground}{%
    \parbox{\dimexpr\linewidth-2\fboxsep-2\fboxrule\relax}{%
      \raggedright
      \hyphenpenalty=10000\exhyphenpenalty=10000\relax
      \textbf{Finding \thefinding:} \emph{#1}\par
    }%
  }%
  \endgroup
  \par\addvspace{0.5\baselineskip}
}

\title{Can AI Agents Detect and Repair Artifact Drift in Network Experiments?}

\author{Tianzhu Zhang}
\authornote{Equal contribution.}
\affiliation{%
  \institution{Nokia Bell Labs}
  \city{Massy}
  \country{France}
}
\email{tianzhu.zhang@nokia-bell-labs.com}

\author{Weichen Tao}
\authornotemark[1]
\affiliation{%
  \institution{Telecom Paris}
  \city{Palaiseau}
  \country{France}
}
\email{weichen.tao@telecom-paris.fr}

\author{Changgang Zheng}
\authornotemark[1]
\affiliation{%
  \institution{Nanjing University}
  \city{Nanjing}
  \country{China}
}
\email{changgangzheng@qq.com}

\author{Yusheng Zheng}
\affiliation{%
  \institution{University of California, Santa Cruz}
  \city{Santa Cruz}
  \state{California}
  \country{USA}
}
\email{yzhen165@ucsc.edu}

\author{Long Chen}
\affiliation{%
  \institution{The University of Hong Kong}
  \city{Hong Kong}
  \country{Hong Kong SAR, China}
}
\email{chenlong.sjtu@gmail.com}

\author{Xiaoyi Fan}
\affiliation{%
  \institution{Tsinghua University}
  \city{Beijing}
  \country{China}
}
\email{fan-xy25@mails.tsinghua.edu.cn}

\author{Meikang Qiu}
\affiliation{%
  \institution{Augusta University}
  \city{Augusta}
  \state{Georgia}
  \country{USA}
}
\email{qiumeikang@gmail.com}

\begin{document}

\begin{abstract}

In recent years, AI agents have evolved into capable assistants that carry out multi-step tasks in digital environments. The network systems community is beginning to explore these capabilities in operational and experimental settings. However, an agent operating in network systems should not be judged solely by whether it completes the immediate task. The experiment record it modifies must also remain trustworthy. We call this property \emph{artifact integrity}: the record's claims must remain supported by the available evidence, confined to the scope established by that evidence, and traceable through the artifacts that encode their support.

To make this property measurable, we introduce \bench, which tests whether AI agents can repair inconsistent records derived from public network-system artifacts while preserving claims that remain supported. The benchmark contains 52 instances with injected inconsistencies ranging from direct contradictions to unstated relations spread across several artifacts. We evaluate 23 agent configurations across three general-purpose AI agent runtimes using deterministic scoring. The average contract pass rate is 65.3\% across 5,980 outputs, but no agent runtime exceeds 30\% when repair requires recovering implicit relations and propagating changes across artifacts. These results reveal a sharp boundary between local correction and complete record-level repair. Therefore, we argue that artifact integrity should become a first-class design and evaluation requirement for AI agents operating on network systems.
\end{abstract}

\maketitle

\section{Introduction}

With the rise of agentic workflows powered by large language models (LLMs), AI agents are beginning to reshape network systems, from operations to experimentation. Network automation has already evolved beyond static procedures toward policy and intent-driven management, closed-loop control, and data-driven optimization~\cite{clemm2022intent}. LLM-powered agents extend this trajectory by interpreting heterogeneous evidence, acting on the network through tools, and revising the artifacts that record their actions and outcomes. Recent work has explored AI agents for network diagnosis, intent-driven management, configuration, and Internet measurement~\cite{wang2024netassistant,wang2025intent,wang2025nika,twabi2026netagentbench,ramanathan2025towards}. Industry roadmaps likewise position AI agents as a key component of future autonomous networking~\cite{newman2025agentic,ericsson2025aiagents}.

This shift changes not only how network tasks are executed, but also how knowledge about the network is assembled and carried forward. An agent may inspect configurations, combine observations from several sources, and revise the experiment record. The record captures what was executed, under which conditions, what was observed, and what the evidence supports. As AI agents take on larger roles in network operations and experimentation, such records may inform subsequent diagnoses, configurations, and operational decisions. The integrity of these records must be preserved whenever downstream tasks depend on them.

\begin{figure*}[!t]
  \centering
  \includegraphics[width=0.95\textwidth]{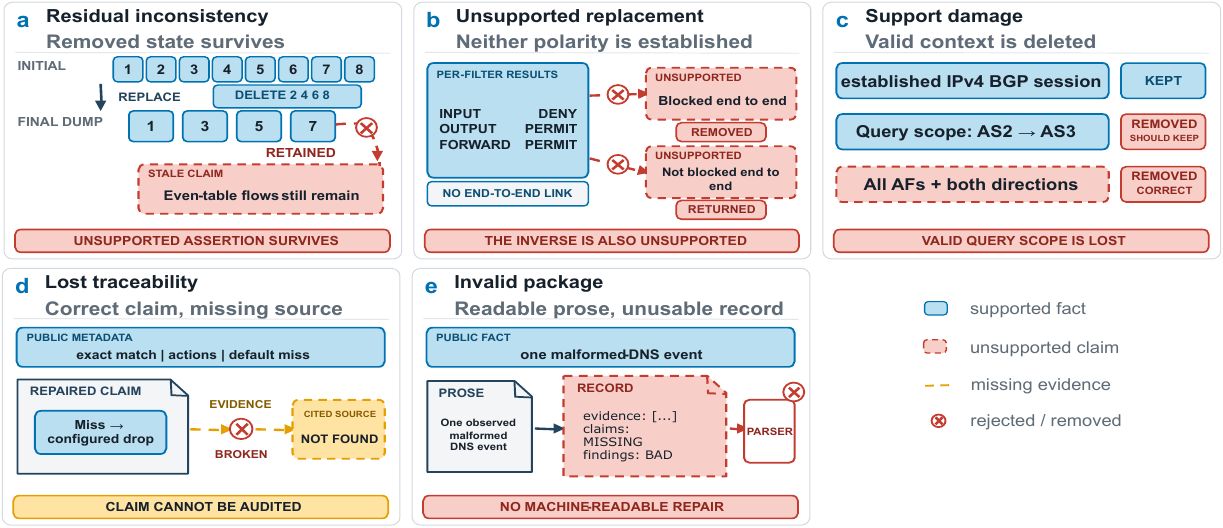}
  \caption{Five contract-visible failure modes in agent-produced repairs. 
  }
  \Description{ (a) Residual inconsistency preserves a stale assertion. (b) Unsupported replacement substitutes an ungrounded inverse. (c) Support damage removes valid context. (d) Lost traceability breaks the evidence path. (e) An invalid package prevents downstream consumption. Five schematics show: removed Open vSwitch flows carried into a stale final claim; opposite end-to-end claims inferred from independent filter outcomes; a valid BGP query scope removed with an overclaim; a corrected P4 claim linked to a missing source; and a malformed Zeek repair rejected by a parser.}
  \label{fig:repair-failure-modes}
  \vspace{-3mm}
\end{figure*}

However, current evaluations remain largely outcome-centered. Determining whether an agent achieves the requested result is necessary, but it does not reveal whether the artifacts it modifies remain faithful to the underlying evidence. We refer to the configurations, logs, reports, and structured claims associated with a network task as its \emph{network experiment package}; together, they form the experiment record. An agent may satisfy the immediate objective while leaving this record inconsistent with, or unsupported by, the available evidence. We call this failure \emph{artifact drift}. Preventing it requires \emph{artifact integrity}: every claim must remain supported by the evidence, bounded by its scope, and traceable to the artifacts that justify it. A valid repair must propagate consistently across the record and leave enough provenance for another reviewer, tool, or agent to reconstruct what changed and why.

Artifact drift becomes more consequential as agents gain greater autonomy. In an assisted task, drift may distort one result. When an agent carries findings from one experiment into the next, however, the record becomes part of the state that guides future actions. An unsupported claim may then shape a new hypothesis, configuration, or measurement, propagating a local inconsistency through an entire line of work. 
Existing benchmarks remain largely outcome-centered~\cite{wang2025nika,twabi2026netagentbench,baek2026artisan,wu2026artifactcopilot}. They do not ask whether the surrounding experiment record remains coherent, evidence-grounded, and supportable after the agent has acted. Artifact integrity thus remains a missing dimension in the evaluation of AI agents for network systems.

We introduce \bench\ to make this missing target measurable. Rather than treating correctness as a collection of file-local checks, \bench\ models artifact integrity through the relations among claims, evidence, scope, provenance, and experimental phase. It realizes this view in 52 source-grounded instances spanning four network task families, each combining controlled drift with valid claims that an agent must preserve. The benchmark further provides a deterministic evaluation protocol that tests whether the agent identifies the broken relation, propagates the correction across all affected artifacts, and leaves behind a complete and traceable record.



\section{How Artifact Repairs Fail}

\bench\ presents each agent with a record containing a controlled integrity violation. Fig.~\ref{fig:repair-failure-modes} summarizes five ways the returned repair can fail: it may leave the seeded violation unresolved or introduce a new substantive or auditability failure. The modes are illustrative and may overlap.

\paragraph{Residual inconsistency:}
A repair may correct the injected inconsistency in one artifact while leaving the same unsupported assertion elsewhere in the experiment record. Fig.~\ref{fig:repair-failure-modes} (a) shows an example in an Open vSwitch experiment~\cite{pfaff2015ovs}: an initial phase installs several flows, a later replacement removes some of them, and the final state contains only the surviving flows. The returned package preserves the individual phase records but still claims that the removed flows remain in the final state. The source artifacts are valid; the stale conclusion remains unresolved.

\paragraph{Unsupported replacement:}
Removing an unsupported conclusion does not justify asserting its opposite. In the Batfish example of Fig.~\ref{fig:repair-failure-modes} (b)~\cite{fogel2015batfish}, the public evidence records how individual filters treat a test flow but does not establish end-to-end reachability. The repair replaces the unsupported claim that the flow is blocked end-to-end with the equally unsupported claim that it is not. The direction of the conclusion changes, but the missing path relation is never established.

\paragraph{Support damage:}
A repair causes support damage when it removes valid information together with an unsupported claim. In Fig.~\ref{fig:repair-failure-modes} (c), Batfish output establishes the state of a particular BGP session, and the query records the direction and address family. The report incorrectly generalizes this result to sessions in all directions and address families. A correct repair would narrow the claim to the session and scope actually tested. Instead, the repair deletes both the unsupported generalization and the valid statement describing what the query covered. The false claim is removed, but the remaining observation loses the scope needed to interpret it.

\paragraph{Lost traceability:}
A correction may be substantively plausible while leaving its support impossible to verify from the experiment record. In the P4 example of Fig.~\ref{fig:repair-failure-modes} (d)~\cite{bosshart2014p4}, the repair narrows an end-to-end forwarding claim to the configured default-miss behavior, but points to a source location that does not exist in the public package. The revised claim may be correct, yet the record no longer provides a valid link between that claim and the evidence intended to support it.

\paragraph{Invalid package:}
A repair may express a plausible correction without returning it in a valid experiment record. In the Zeek-derived case of Fig.~\ref{fig:repair-failure-modes} (e)~\cite{paxson1999bro}, the agent provides a readable explanation but no parseable collection of repaired claims. The returned package thus does not establish which claims remain, how they are classified, or which evidence supports them. Restoring artifact integrity requires not only coherent corrections but also a valid representation in which those corrections can be identified and verified.

These failure modes are distinct but may coexist in one output. A malformed package may also retain a stale claim, and a coordinated repair may simultaneously introduce support damage. Together, they reveal two requirements. First, the returned record must restore the substantive relations among claims, evidence, scope, provenance, and experimental phase. Second, those relations must remain explicit enough for another tool or agent to audit and reuse. The next section formalizes the first requirement, and Section~\ref{sec:design} turns both into a deterministic benchmark contract.


\section{Claim--Evidence--Scope Model}
\label{sec:model}

The failures above arise when an agent changes individual artifacts without restoring the relations that support the experiment record. Following research-object models of scientific artifacts~\cite{bechhofer2013researchobject,soilandreyes2022rocrate}, we represent a network experiment package as a graph of claims, evidence, scope constraints, and provenance or phase relations. A claim states what the experiment establishes, evidence records what was configured or observed, and scope specifies the conditions under which the claim holds. Artifact integrity requires the public package to establish a valid path from evidence to each claim.

A support path may be direct, as when a value is read from a measurement, or relational, as when a claim depends on joining observations, binding metadata objects, or following state across phases. This view draws on database and general provenance models~\cite{buneman2001whywhere,cheney2009provenance,moreau2013provdm}. A path is valid only if every required connection is established by the public artifacts. Matching identifiers do not by themselves justify a cross-trace join, coexisting objects do not imply a binding, and observations from different phases cannot be combined without respecting their order and provenance.

For a claim \(c\), let \(\mathcal{P}_c^{+}\) denote its valid public support paths, and let \(\mathrm{ctx}(p)\) denote the conditions under which path \(p\) is valid. A claim presented as established satisfies artifact integrity only if
\[
\exists p \in \mathcal{P}_c^{+}
\quad\text{such that}\quad
\mathrm{scope}_{\mathrm{eff}}(c)
\subseteq
\mathrm{ctx}(p).
\]
Here, \(\mathrm{scope}_{\mathrm{eff}}(c)\) is the scope asserted after applying explicit and consistent limitations. A limitation may narrow a claim, but it does not supply positive evidence or silently correct broader prose.

We label a claim \textsc{observed} when it follows directly from reported results, \textsc{supported} when other public artifacts establish it, and \textsc{unsupported} when no valid support path exists. An unsupported claim may remain only if it is explicitly labeled as such and nowhere else presented as established. A valid repair must propagate the correction across the record, preserve claims that remain supported, and avoid fabricating missing evidence or relations. Section~\ref{sec:design} turns these requirements into a benchmark contract and adds the parseability and traceability conditions needed for machine auditability.


\section{\bench\ Design}
\label{sec:design}
\begin{figure*}[!tb]
\centering
\includegraphics[width=0.9\textwidth]{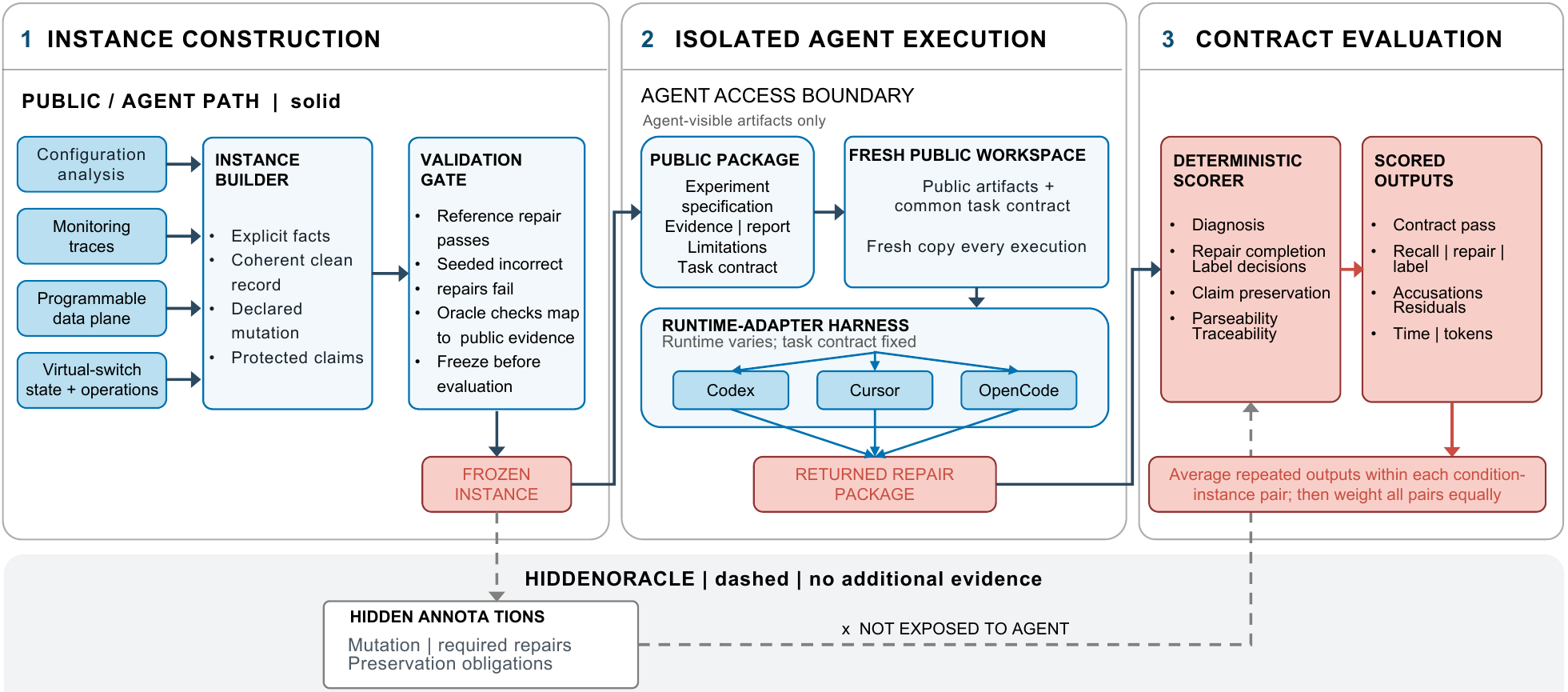}
\caption{\bench\ construction and evaluation workflow. Public source facts are assembled into a coherent package; a declared mutation is injected without changing those facts, and the instance is validated and frozen. Every execution gives the selected runtime a fresh copy of only the public package and task contract. Hidden annotations add no evidence, bypass the agent, and meet the returned repair only at the deterministic scorer, which checks diagnosis, repair completion, labels, preservation, parseability, and traceability. Solid paths are public or agent-visible; dashed gray paths are hidden.}
\Description{A three-stage, two-lane workflow. Four public network-system source families feed an instance builder and validation gate. A frozen instance splits into a solid public branch that crosses the agent-access boundary into a fresh workspace and runtime-adapter harness, and a dashed hidden-oracle branch that bypasses the agent. The returned repair and hidden annotations meet only at a deterministic scorer, which emits contract and component metrics before aggregation across five repetitions.}
\label{fig:benchmark-workflow}
\vspace{-2mm}
\end{figure*}

\bench\ turns the claim--evidence--scope model into a self-contained repair task over network-system experiment records. The agent receives a frozen public package and task contract, but no hidden oracle information. It must identify the violated relation, propagate the correction across affected artifacts, preserve claims that remain supported, and return parseable and traceable artifacts. Figure~\ref{fig:benchmark-workflow} summarizes construction, execution, and scoring.

\subsection{Instance Construction}
\bench\ contains 52 instances derived from public Batfish~\cite{fogel2015batfish}, Zeek~\cite{paxson1999bro}, P4C~\cite{p4lang2026p4c}, and Open vSwitch~\cite{pfaff2015ovs} repositories. Their tests, traces, compiler outputs, expected results, and phase-ordered state provide stable evidence anchors. Each injected inconsistency breaks a relation grounded in source-project semantics, such as scope containment, trace provenance, metadata binding, or temporal ordering. The public task contract declares permitted commands, required outputs, and citation rules; every oracle judgment is grounded in public evidence.
The resulting instances differ not only in network domain but also in how the drift is planted and what a successful repair must reconstruct. Specifically, some contain an explicit contradiction in a single artifact, while others require the agent to recover a relation across provenance, scope, metadata, or experimental phases. We organize these differences into five structural tiers, summarized in Table~\ref{tab:structure-tiers}.
The tiers progress from direct correction (Tiers~1--2) to explicit relational repair (Tier~3), coordinated propagation (Tier~4), and inferred relations (Tier~5).

\begin{table}[!tb]
\caption{Structural tiers of artifact drift.}
\label{tab:structure-tiers}
\scriptsize
\begin{tabularx}{\columnwidth}
{@{}p{0.10\columnwidth}Y@{}}
\toprule
Tier & What the repair requires \\
\midrule
1 & Correct one directly exposed contradiction in a metric, scope, command, label, or limitation. \\
2 & Correct several exposed contradictions while preserving claims that remain supported. \\
3 & Follow an explicit relation or scope constraint across rows, filters, phases, identifiers, or attachments. \\
4 & Propagate one correction across several affected artifacts. \\
5 & Infer an unstated relation across tables, source contexts, or control and runtime artifacts. \\
\bottomrule
\end{tabularx}
\end{table}


\subsection{Machine-Auditable Scoring}

A structured finding matches a violation when it identifies the expected claim, field, artifact scope, and observed-to-corrected value relation. The repaired experiment report is checked for repair completeness and preservation. The human-readable audit report must be present, but its prose is not scored.
Let $M$ denote the scored violations, $R$ their corresponding repair obligations, and $L$ the required post-repair label decisions. The scorer also tracks supported claims that must remain intact. We report the following metrics:
\[
\begin{aligned}
\mathrm{recall} &=
\frac{\text{scored violations credited as detected}}{|M|},\\
\mathrm{repair} &=
\frac{\text{satisfied repair obligations}}{|R|},\\
\mathrm{label} &=
\frac{\text{correct label decisions}}{|L|},\\
\mathrm{accus.} &=
\text{protected-claim accusation events},\\
\mathrm{residual} &=
\text{remaining contract violations}.
\end{aligned}
\]
An accusation event occurs when a confirmed structured finding attacks a protected claim or the returned claim-support map removes or incorrectly relabels it. Other supported-content preservation failures contribute to residual violations but not necessarily to \emph{Accus.} Residual violations also include unresolved repairs, schema errors, and invalid evidence references. Limitation-drift mutations and their associated repairs are removed from the per-output \emph{Recall} and \emph{Repair} denominators. For every reported group, we average the output-level metrics over the applicable instances and agent configurations within each repetition, then report the mean and sample standard deviation across the five repetitions.

These metrics determine a single contract-level decision. An agent output passes only if every scored violation is identified, every required repair is completed, all protected claims are preserved, no residual contract violation remains, and all prescribed artifacts are parseable and traceable. Traceability requires valid references to public evidence and the exact preservation of required public claim identifiers.



\begin{table*}[!tb]
\caption{Average performance metrics across agent runtimes and LLM backends. Entries report the mean $\pm$ sample standard deviation across five repetitions, each averaged over all 52 instances. 
}
\label{tab:agent-configuration-results}
\centering
\scriptsize
\setlength{\tabcolsep}{0.8pt}
\begin{tabular*}{\textwidth}{@{\extracolsep{\fill}}llrrrrrr@{}}
\toprule
\multirow{2}{*}{Runtime} & \multirow{2}{*}{Backend / setting} &
\multicolumn{4}{c}{Artifact integrity (\% $\uparrow$)} &
\multicolumn{2}{c}{Violations ($\downarrow$)} \\
\cmidrule(lr){3-6}\cmidrule(l){7-8}
 & & Contract & Recall & Repair & Label & Accus. & Resid. \\
\midrule
\multirow{9}{*}{\textbf{Codex}} & GPT-5.3 Codex Spark / high & \contractplain{66.9}\metricstd{3.7} & \metricmeanstd{74.6}{4.2} & \metricmeanstd{78.7}{3.2} & \metricmeanstd{94.3}{1.8} & \metricmeanstd{0.000}{0.000} & \metricmeanstd{0.642}{0.057} \\
 & GPT-5.4 mini / medium & \contractlow{68.1}\metricstd{4.2} & \metricmeanstd{80.2}{2.2} & \metricmeanstd{83.1}{2.5} & \metricmeanstd{98.5}{0.8} & \metricmeanstd{0.012}{0.017} & \metricmeanstd{0.669}{0.096} \\
 & GPT-5.4 / medium & \contractmid{71.2}\metricstd{1.9} & \metricmeanstd{85.8}{2.8} & \metricmeanstd{85.4}{0.8} & \metricmeanstd{99.2}{1.1} & \metricmeanstd{0.008}{0.017} & \metricmeanstd{0.569}{0.070} \\
 & GPT-5.5 / low & \contractmid{73.1}\metricstd{2.7} & \metricmeanstd{84.6}{1.2} & \metricmeanstd{86.3}{1.6} & \metricmeanstd{99.1}{1.0} & \metricmeanstd{0.004}{0.009} & \metricmeanstd{0.523}{0.067} \\
 & GPT-5.5 / medium & \contractmid{76.5}\metricstd{1.6} & \metricmeanstd{86.9}{2.2} & \metricmeanstd{88.7}{0.8} & \metricmeanstd{99.6}{0.6} & \metricmeanstd{0.012}{0.017} & \metricmeanstd{0.446}{0.042} \\
 & GPT-5.5 / high & \contractmid{76.9}\metricstd{2.4} & \metricmeanstd{88.5}{1.8} & \metricmeanstd{88.5}{1.2} & \metricmeanstd{99.4}{0.6} & \metricmeanstd{0.012}{0.011} & \metricmeanstd{0.442}{0.045} \\
 & GPT-5.5 / xhigh & \contracttop{81.9}\metricstd{2.9} & \metricmeanstd{90.8}{2.0} & \metricmeanstd{90.6}{1.9} & \metricmeanstd{99.8}{0.3} & \metricmeanstd{0.000}{0.000} & \metricmeanstd{0.346}{0.090} \\
 & GPT-5.6 Luna / high & \contractmid{75.0}\metricstd{1.4} & \metricmeanstd{90.2}{1.6} & \metricmeanstd{86.2}{1.1} & \metricmeanstd{97.2}{1.7} & \metricmeanstd{0.000}{0.000} & \metricmeanstd{0.454}{0.046} \\
 & GPT-5.6 Sol / high & \contractmid{78.5}\metricstd{1.6} & \metricmeanstd{91.2}{2.1} & \metricmeanstd{88.8}{0.9} & \metricmeanstd{99.7}{0.3} & \metricmeanstd{0.000}{0.000} & \metricmeanstd{0.392}{0.083} \\
\midrule
\multirow{8}{*}{\textbf{Cursor}} & Claude Opus 4.8  & \contracttop{80.8}\metricstd{1.9} & \metricmeanstd{92.9}{1.7} & \metricmeanstd{91.0}{1.1} & \metricmeanstd{100.0}{0.0} & \metricmeanstd{0.000}{0.000} & \metricmeanstd{0.350}{0.029} \\
 & Claude Sonnet 5  & \contractmid{80.0}\metricstd{2.2} & \metricmeanstd{90.8}{2.0} & \metricmeanstd{89.6}{1.7} & \metricmeanstd{98.9}{0.9} & \metricmeanstd{0.000}{0.000} & \metricmeanstd{0.358}{0.092} \\
 & Composer 2.5 & \contractmid{75.4}\metricstd{2.1} & \metricmeanstd{92.5}{1.4} & \metricmeanstd{88.3}{1.4} & \metricmeanstd{99.6}{0.4} & \metricmeanstd{0.000}{0.000} & \metricmeanstd{0.504}{0.032} \\
 & Gemini 3.1 Pro & \contractplain{66.9}\metricstd{3.7} & \metricmeanstd{79.6}{1.3} & \metricmeanstd{81.9}{1.3} & \metricmeanstd{98.3}{1.2} & \metricmeanstd{0.008}{0.017} & \metricmeanstd{0.681}{0.050} \\
 & Gemini 3.5 Flash & \contractmid{70.0}\metricstd{3.5} & \metricmeanstd{87.9}{3.7} & \metricmeanstd{84.2}{1.5} & \metricmeanstd{97.8}{0.9} & \metricmeanstd{0.000}{0.000} & \metricmeanstd{0.608}{0.097} \\
 & GPT-5.4  & \contractmid{72.7}\metricstd{0.9} & \metricmeanstd{86.7}{2.1} & \metricmeanstd{86.5}{1.0} & \metricmeanstd{99.2}{0.7} & \metricmeanstd{0.019}{0.024} & \metricmeanstd{0.531}{0.029} \\
 & GPT-5.4 nano  & \contractplain{45.4}\metricstd{8.2} & \metricmeanstd{53.1}{6.4} & \metricmeanstd{65.6}{5.4} & \metricmeanstd{79.4}{3.9} & \metricmeanstd{0.081}{0.032} & \metricmeanstd{1.269}{0.296} \\
 & Grok 4.5  & \contractmid{78.5}\metricstd{0.9} & \metricmeanstd{95.4}{0.8} & \metricmeanstd{89.4}{0.7} & \metricmeanstd{99.9}{0.2} & \metricmeanstd{0.000}{0.000} & \metricmeanstd{0.400}{0.025} \\
\midrule
\multirow{6}{*}{\textbf{OpenCode}} & DiffusionGemma & \contractplain{48.1}\metricstd{6.5} & \metricmeanstd{55.8}{4.8} & \metricmeanstd{62.9}{3.9} & \metricmeanstd{77.7}{5.8} & \metricmeanstd{0.054}{0.039} & \metricmeanstd{1.462}{0.224} \\
 & DeepSeek V4 Flash & \contracttop{72.3}\metricstd{5.5} & \metricmeanstd{86.7}{3.9} & \metricmeanstd{83.1}{4.1} & \metricmeanstd{93.3}{2.6} & \metricmeanstd{0.042}{0.029} & \metricmeanstd{0.500}{0.091} \\
 & MiniMax M2.5 & \contractlow{66.2}\metricstd{10.0} & \metricmeanstd{80.2}{10.6} & \metricmeanstd{79.4}{11.2} & \metricmeanstd{90.7}{12.9} & \metricmeanstd{0.031}{0.029} & \metricmeanstd{0.919}{0.705} \\
 & Qwen3.6-35B & \contractlow{60.8}\metricstd{5.0} & \metricmeanstd{75.8}{2.7} & \metricmeanstd{73.5}{2.5} & \metricmeanstd{84.2}{2.6} & \metricmeanstd{0.004}{0.009} & \metricmeanstd{0.858}{0.193} \\
 & Qwen3 8B & \contractplain{1.2}\metricstd{1.7} & \metricmeanstd{1.7}{1.4} & \metricmeanstd{2.9}{1.8} & \metricmeanstd{3.8}{2.2} & \metricmeanstd{0.012}{0.026} & \metricmeanstd{3.219}{0.623} \\
 & Qwen3 14B & \contractplain{15.4}\metricstd{4.5} & \metricmeanstd{24.4}{4.2} & \metricmeanstd{31.5}{4.9} & \metricmeanstd{39.4}{6.8} & \metricmeanstd{0.008}{0.017} & \metricmeanstd{1.765}{0.177} \\
\bottomrule
\end{tabular*}
\end{table*}


\section{Evaluation}

We organize the evaluation around three research questions:
\begin{itemize}
    \item \textbf{RQ1:} How effectively can current general-purpose AI agents restore artifact integrity?
    \item \textbf{RQ2:} How does repair reliability change as tasks impose increasingly complex relations across artifacts?
    \item \textbf{RQ3:} In what ways do current artifact-integrity repairs remain unreliable?
\end{itemize}

\paragraph{Experimental setup:}
For this study, an \emph{AI agent} is the deployed combination of a backend LLM and an \emph{agent runtime}.
The LLM provides the underlying reasoning and generation capabilities, while the runtime manages context, exposes tools and files, and carries out edits in the workspace.
Each \emph{agent configuration} fixes the runtime, backend LLM, and exposed inference settings.

We construct 23 agent configurations using Codex CLI, Cursor Agent, and OpenCode to inspect and edit \seqsplit{self-contained} workspaces that directly match \bench's interface.
This combination lets us vary model capability and inference settings without changing the benchmark or package format.
The 23 agent configurations and 52 instances define 1,196 configuration--instance pairs. Each agent configuration is executed five times on every instance, yielding 5,980 evaluable outputs\footnote{We use Codex CLI versions 0.142.3--0.144.5 with GPT-5.3 Codex Spark, GPT-5.4, GPT-5.4 mini, GPT-5.5, GPT-5.6 Luna, and GPT-5.6 Sol. 
Cursor Agent build \texttt{2026.07.01-41b2de7} provides Claude Opus~4.8, Composer~2.5, Gemini~3.5 Flash, GPT-5.4, and GPT-5.4 nano; build \texttt{2026.07.23-e383d2b} provides Cursor Grok~4.5, Gemini~3.1 Pro, and Claude Sonnet~5.
OpenCode~1.17.15 runs Qwen3 8B and Qwen3 14B locally through Ollama and connects to freeinference~\cite{freeinference} for MiniMax M2.5, DiffusionGemma, DeepSeek V4 Flash, and Qwen3.6-35B}.
This work focuses on general-purpose agents.
Network-specialized agents contain built-in telemetry, simulators, APIs, and task-specific state~\cite{wang2025nika,twabi2026netagentbench,wang2025intent,ramanathan2025towards}.
Testing them would make it difficult to distinguish whether performance came purely from artifact-integrity reasoning or from additional support by the specialized environment.


\subsection{RQ1: Overall Effectiveness}

Overall, 3,904 of the 5,980 outputs pass the contract (65.3\%), with 74.2\% for Codex, 71.2\% for Cursor, and 44.0\% for OpenCode (aggregated from Table~\ref{tab:agent-configuration-results}). For GPT-5.5 under Codex, contract pass rates rise from 73.1\% at low effort to 76.5\%, 76.9\%, and 81.9\% at medium, high, and xhigh effort. The added GPT-5.3 Codex Spark, GPT-5.6 Luna, and GPT-5.6 Sol configurations attain 66.9\%, 75.0\%, and 78.5\%, respectively. The added Cursor Grok~4.5/high, Gemini~3.1 Pro, and Claude Sonnet~5/high configurations attain 78.5\%, 66.9\%, and 80.0\%, respectively. Configuration-level rates span 66.9--81.9\% within Codex and 45.4--80.8\% within Cursor.
The six OpenCode configurations span 1.2--72.3\%, with DeepSeek V4 Flash attaining the highest OpenCode rate.

\findingbox{The evaluated general-purpose AI agents do not reliably restore artifact integrity in this suite.
}

\subsection{RQ2: Structural Complexity}
Contract pass rates decline across the structural tiers in this suite. From Tier~1 to Tier~5, they fall from 98.9\% to 26.3\% for Codex, from 93.8\% to 28.4\% for Cursor, and from 60.8\% to 17.0\% for OpenCode (Table~\ref{tab:tier-contract}).
All three runtimes exhibit lower pass rates in more challenging structural tiers that contain coordinated or inferred cross-artifact relations.
\findingbox{Across all three runtimes, contract pass rates decline as repairs move from directly exposed contradictions (Tiers~1--2) to following, propagating, or inferring relations across artifacts (Tiers~3--5).}

\begin{table}[!tb]
\caption{Contract pass rate (\%) by structural tier (mean $\pm$ sample standard deviation across five repetitions).}
\label{tab:tier-contract}
\centering
\scriptsize
\begin{tabular*}{\columnwidth}{@{\extracolsep{\fill}}lrrrrr@{}}
\toprule
Runtime & 1 & 2 & 3 & 4 & 5 \\
\midrule
Codex & \metricmeanstd{98.9}{0.4} & \metricmeanstd{97.8}{2.2} & \metricmeanstd{78.0}{3.1} & \metricmeanstd{49.3}{5.8} & \metricmeanstd{26.3}{2.3} \\
Cursor & \metricmeanstd{93.8}{1.5} & \metricmeanstd{86.5}{5.5} & \metricmeanstd{80.2}{2.1} & \metricmeanstd{40.0}{3.1} & \metricmeanstd{28.4}{2.3} \\
OpenCode & \metricmeanstd{60.8}{4.4} & \metricmeanstd{49.3}{3.7} & \metricmeanstd{47.3}{2.9} & \metricmeanstd{23.3}{4.7} & \metricmeanstd{17.0}{4.5} \\
\bottomrule
\end{tabular*}
\vspace{-2.5mm}
\end{table}

\subsection{RQ3: Reliability Limits}

\paragraph{Partial contract completion:}
This partial completion appears at every structural tier. For every runtime--tier combination, mean recall, repair, and label rates exceed the contract pass rate, with higher structural tiers exhibiting wider gaps.
Tier~5 provides the clearest illustration. Label accuracy exceeds the contract pass rate for every runtime.
Mean label accuracies are 96.2\%, 93.4\%, and 60.2\% for Codex, Cursor, and OpenCode, while their contract pass rates are 26.3\%, 28.4\%, and 17.0\%.
The corresponding recall rates are 51.1\%, 57.0\%, and 33.9\%, and repair rates are 59.6\%, 58.1\%, and 35.0\%.
As the contract pass requires that every obligation be jointly fulfilled, these rates show that agents often complete only part of the repair, but do not reveal which unmet obligation most often prevents a full pass.
The hardest instance-level outcomes likewise span the later tiers.
Two cross-artifact instances receive no contract pass in any of 115 outputs per instance: one Tier~4 scope-propagation task and one Tier~5 join task drawn from programmable data-plane metadata.
Every returned package for these instances violates at least one declared contract obligation.

\paragraph{Observed failure flags:}
One failed output can violate several contract obligations and thus contribute to multiple failure categories, so the percentages below do not sum to 100\%.
Among the 2,076 output-level failures, 68.5\% leave at least one required diagnosis or repair incomplete, 47.9\% omit required report information, 33.6\% fail at least one supported-content preservation check, and 22.6\% return invalid structured output.
Required-artifact omission appears in 7.1\%, and 3.9\% retain the injected unsupported conclusion.

\paragraph{Run-to-run instability:}
Among the 1,196 \seqsplit{configuration-instance} pairs, 331 (27.7\%) contain both passing and failing outputs.
Mixed outcomes occur in 19.9\% of Codex pairs, 25.0\% of Cursor pairs, and 42.9\% of OpenCode pairs.
For the smallest Cursor backend, the first-execution pass rate is 59.6\%, compared with 45.4\% across five executions.

\findingbox{The evaluated agents often satisfy individual repair obligations without completing the full artifact integrity contract, and repeated executions can change whether the repair passes.}


\section{Related Work}

Several works explore applying LLMs and AI agents to network systems. \textsc{NetConfEval}~\cite{wang2024netconfeval} and VPP~\cite{mondal2023vpp} evaluate configuration generation and correction, while \textsc{NetAssistant}~\cite{wang2024netassistant}, \textsc{LLexus}~\cite{lascasas2024llexus}, \textsc{OSS-GPT}~\cite{mekrache2025oss}, \textsc{Confucius}~\cite{wang2025intent}, and \textsc{ArachNet}~\cite{ramanathan2025towards} support multi-step diagnosis, operations, and measurement workflows. Dedicated AI agent benchmarks make these tasks testable and reproducible. \textsc{Cornetto}~\cite{protogeros2026cornetto} checks configuration repairs and regressions, \textsc{NIKA}~\cite{wang2025nika} replays dynamic incidents, and \textsc{NetAgentBench} \cite{twabi2026netagentbench} evaluates multi-turn configuration through a deterministic state machine. In these works, the evaluation unit remains the immediate configuration, incident, or workflow outcome. \bench\ evaluates whether a repair restores the relations within an experiment record while preserving claims that remain supported.

Outside the networking domain, \textsc{SWE-bench} evaluates whether repository edits resolve a software issue, while \textsc{FEVER} tests whether textual evidence supports or refutes a claim~\cite{jimenez2023swebench,thorne2018fever}. \textsc{Artisan} evaluates scripts that reproduce published results, and \textsc{ArtifactCopilot} automates environment setup and recovery during artifact execution~\cite{baek2026artisan,wu2026artifactcopilot}. These tasks evaluate issue resolution, claim classification, result reproduction, and successful execution, respectively. \bench\ instead evaluates whether a repair leaves an entire network experiment record coherent, traceable, and free of damage to claims that remain supported.


\section{Conclusion}

As AI agents begin to enter network systems, the artifacts they modify become part of the record from which results are interpreted, verified, and reused. A repair can appear locally plausible while leaving that record internally inconsistent or no longer traceable to its evidence. \bench\ makes this record-level failure measurable through 52 source-grounded instances that require agents to identify broken claim--evidence relations, propagate corrections across affected artifacts, and preserve supported claims.


Our evaluation reveals that strong general-purpose AI agents can handle direct contradictions well, but struggle when repair depends on implicit relations or coordinated changes across artifacts. Local correctness does not ensure a complete and auditable record of the experiment. These results make artifact integrity a first-class requirement for AI agents operating on network systems. Reliable AI agents for network systems must preserve both the record's substantive support and the explicit structure that later tools and agents need to inspect and reuse it.

The controlled violations probe repair capability, not the prevalence or natural distribution of integrity failures in real network workflows. Future work should study naturally occurring cases, independent oracle review, more permissive semantic evaluation, and network-specialized agents.

\bibliographystyle{ACM-Reference-Format}
\bibliography{references}

\end{document}